\documentclass[sigconf,nonacm]{acmart}
\setcopyright{none}
\renewcommand\footnotetextcopyrightpermission[1]{}
\usepackage{amsmath,amsfonts}
\usepackage{algorithmic}
\usepackage{graphicx}
\usepackage{textcomp}
\usepackage{xcolor}
\usepackage{adjustbox}
\usepackage{booktabs}
\usepackage{pifont}
\usepackage{enumitem}
\usepackage{amsmath}
\usepackage{extarrows}
\usepackage{listings}
\usepackage{makecell}
\usepackage{booktabs}
\usepackage{multirow}
\usepackage{comment}
\usepackage{wrapfig}

\usepackage[pdftex,svgnames,table]{xcolor}
\makeatletter
\@ifpackageloaded{hyperref}{
  \AtBeginDocument{\hypersetup{
    colorlinks=true,
    linkcolor=blue,    
    citecolor=teal,    
    urlcolor=magenta,  
    breaklinks=true,
  }}
}{
  \usepackage[
    colorlinks=true,
    linkcolor=blue,    
    citecolor=teal,    
    urlcolor=magenta,  
    breaklinks=true,
  ]{hyperref}
}
\makeatother

\makeatletter
\def\NAT@pd@first#1{\expandafter\NAT@pd@first@i\detokenize{#1}\NAT@pd@nil}
\def\NAT@pd@first@i#1#2\NAT@pd@nil{\ifcat A\noexpand#1}
\def\NAT@parse@date#1#2#3#4#5#6@@{%
  \NAT@pd@first{#1}\def\NAT@year{}\def\NAT@exlab{#1}\else
  \NAT@pd@first{#2}\def\NAT@year{#1}\def\NAT@exlab{#2}\else
  \NAT@pd@first{#3}\def\NAT@year{#1#2}\def\NAT@exlab{#3}\else
  \NAT@pd@first{#4}\def\NAT@year{#1#2#3}\def\NAT@exlab{#4}\else
    \def\NAT@year{#1#2#3#4}\def\NAT@exlab{{#5}}\fi\fi\fi\fi}
\makeatother

\graphicspath{{images/}}
\DeclareGraphicsExtensions{.pdf,.jpeg,.png}

\newcommand{\eg}{{e.g.,}}
\newcommand{\ie}{{i.e.,}}

\newcommand{\etal}{{\it et al.}}

\usepackage{siunitx}

\usepackage{xspace}

\renewcommand{\paragraph}[1]{\vspace{0.03in}\noindent\textbf{#1}}

\makeatletter
\@ifundefined{ACM@origparagraph}{}{\let\ACM@origparagraph\paragraph}
\makeatother

\def\Snospace~{\S{}}

\definecolor{codegray}{rgb}{0.5,0.5,0.5}
\definecolor{codepurple}{rgb}{0.58,0,0.82}
\definecolor{backcolour}{rgb}{0.95,0.95,0.92}

\lstdefinestyle{CStyle}{
  language=C,
  sensitive=true,                     
  basicstyle=\ttfamily\footnotesize,
  keywordstyle=\color{blue}\bfseries, 
  commentstyle=\color{codegray}\itshape,
  stringstyle=\color{codepurple},
  numbers=left,
  numberstyle=\tiny\color{codegray},
  numbersep=5pt,
  breaklines=true,
  keepspaces=true,
  showstringspaces=false,
  showspaces=false,
  showtabs=false,
  tabsize=4,
  alsoletter={_},                     
  alsodigit={0123456789},             
  keywordstyle=[2]\color{teal}\bfseries,
  morekeywords=[2]{uint8_t,uint16_t,uint32_t,bool},
}

\begin{document}

\title{From Temporary Access to Persistent Surveillance:
Why \textsc{Matter} Matters in Smart Homes}

\author{Nicolas Nino}
\affiliation{%
  \institution{University of Georgia}
  \city{Athens}
  \state{Georgia}
  \country{USA}
}

\author{Timothy Minning}
\affiliation{%
  \institution{University of Georgia}
  \city{Athens}
  \state{Georgia}
  \country{USA}
}

\author{Le Guan}
\affiliation{%
  \institution{University of Georgia}
  \city{Athens}
  \state{Georgia}
  \country{USA}
}

\renewcommand{\shortauthors}{Nino et al.}

\begin{abstract}
Matter aims to unify smart home ecosystems through an open, interoperable, and
secure standard, relying on cryptographic mechanisms for device authentication
and data confidentiality. However, its openness also exposes protocol details,
credential structures, and implementation characteristics to adversaries.

We show that an attacker with temporary physical access can exploit design and
implementation flaws to extract credentials and impersonate devices and
controllers. These replicas integrate seamlessly into the fabric, enabling
persistent surveillance and control even after the attacker departs. Notably,
the attack is vendor-agnostic and requires no device-specific reverse
engineering, as long as the device is not physically secured. We validate its
practicality through a proof-of-concept on a simulated smart home with both
commercial devices and development boards. Our findings uncover previously
underestimated attack surfaces in Matter, particularly under physical access
scenarios, and motivate four targeted mitigation strategies.
\end{abstract}



\maketitle


\section{Introduction}

Smart home technologies have rapidly evolved, offering convenience, automation, and energy efficiency
through interconnected IoT devices and central hubs. 
Modern smart homes allow users to remotely control lighting, locks, sensors, and appliances, and 
have become increasingly integrated into everyday life. 
As adoption grows, driven by falling device costs and broader Internet connectivity, 
smart homes are no longer exclusive to tech-savvy users or luxury environments,
but are now commonplace in average households.
However, this growing ubiquity has also introduced a broader attack surface. 
Many smart devices operate continuously, store sensitive data, and interface with both local and cloud-based systems. 
Compounding these security risks is the fragmented nature of the smart home ecosystem, 
where devices from different vendors are often incompatible and offer limited support for interoperability. 
This fragmentation not only hinders the enforcement of uniform security policies, 
exposing systems to configuration errors and integration gaps~\cite{iotmediator, iotmon}, 
but also places practical burdens on users, 
who may be forced to install multiple companion apps, 
manage several hubs, or remain confined to a single vendor’s ecosystem to maintain device compatibility.

To address the security and interoperability challenges posed by the fragmented smart home ecosystem, 
the \textit{Connectivity Standards Alliance (CSA)} introduced the Matter standard~\cite{matter}. 
Matter seeks to unify smart home connectivity through an open, cross-platform protocol, 
backed by major industry stakeholders~\cite{matter-adopters, matter-participants}. 
By adopting the philosophy of security-through-openness,
the Matter specification enables broad scrutiny that benefits manufacturers, developers, and end users.
It mandates strong security mechanisms such as end-to-end encryption and device attestation, to provide consumers with a secure, reliable and interoperable smart home experience.

In this paper, we argue that Matter's design, rooted in transparency and vendor neutrality, 
introduces new risks associated with a previously underestimated attack surface. 
Specifically, when Matter-compliant hardware is deployed in shared environments such as short-term rentals, 
co-living spaces, or hospitality settings~\cite{amazon-iot, august-iot, sm-gadgets-iot, practices-iot, benefits-iot, airbnb-hosts}, 

the conventional assumptions of device ownership, trust, and physical control no longer hold. 
It becomes imperative to reconsider how such hardware can withstand this emerging threat model.

Through a systematic analysis of the Matter protocol and its reference implementation, 
we uncover seven security flaws: three rooted in design and four in implementation (\autoref{sec:flaws}). 
Exploitation of these flaws enables an adversary to:
(1) identify a set of long-lived credentials essential to the cryptographic protection of device identity (\autoref{sec:creidentify});
(2) collect, search for, and derive these credentials from memory or flash dumps of the target hardware, 
and through network eavesdropping (\autoref{sec:credcollection});
(3) replicate the victim hardware by instantiating a firmware template with the extracted credentials; and 
(4) execute stealthy attacks against victim smart homes, such as unauthorized device control 
and falsified data injection without being detected (\autoref{sec:spying}).
We develop a proof-of-concept prototype to validate our findings (\autoref{sec:implementation}). 
Experimental evaluation using both open-source and commercial Matter
controllers, commercial Matter-enabled end devices, and development boards
(Espressif ESP32-S3 and Nordic nRF5340)
confirms that the identified threats are practical and real
(\autoref{Experiment}).

Notably, while the implementation-level vulnerabilities
were initially identified in Matter's reference implementation~\cite{chiptool}, 
they are also present in commercial products, suggesting likely code reuse across vendors.

By exploiting these vulnerabilities, an attacker with only brief physical access
to a victim's home can implant a single covert device into the Matter fabric
(\ie~the Matter network), gaining persistent, fabric-wide visibility and control
without disrupting normal operation. This threat is practical since realistic
access opportunities, such as those granted to a controlling partner, landlord,
Airbnb guest, or repair visitor, provide sufficient unsupervised time to conduct
the attack. Once embedded, the covert device automatically inherits the
capabilities of existing devices, allowing the attacker to observe
controller–device interactions, access detailed sensor readings, and manipulate
devices that victim users believe they legitimately own. A natural alternative might be
to deploy a standalone, non-Matter surveillance device, such as a hidden camera
or microphone. However, such out-of-fabric devices can access only what their
own hardware directly captures. To approach the breadth of information available
within the Matter fabric, an attacker would need to install multiple devices
equipped with different sensor types---cameras, microphones, motion detectors,
contact sensors, and more---dramatically increasing both operational cost and
the risk of detection. In contrast, our single in-fabric device ``senses''
through the victim’s own Matter devices, achieving comprehensive, fabric-wide
monitoring and manipulation that out-of-fabric approaches cannot match.

While our findings expose critical vulnerabilities in the current Matter threat model, 
they also reveal clear avenues for remediation. 
We contend that enhancing Matter’s resilience against physical and transient adversaries 
does not necessitate compromising its open and transparent design philosophy. 
Rather, carefully designed countermeasures can be integrated to reinforce 
security while preserving interoperability and usability. 
In~\autoref{Prevention}, we present four such countermeasures, 
each designed to address a distinct subset of flaws revealed through our analysis.

\noindent In summary, our main contributions are:
\begin{itemize}[leftmargin=*]
    \item We present a comprehensive security evaluation of the Matter standard
    in the context of shared environments, highlighting the mismatch between the
    commonly believed trust model and emerging real-world deployment realities.

    \item We identify seven vulnerabilities (three at design level and four at
    implementation level) through systematic analysis of the Matter
    specification and its reference implementation.
    
    \item We demonstrate the real-world exploitability of these flaws through
     testing on a diverse set of commercial Matter products and development
     kits.

    \item We propose four countermeasure categories to mitigate the identified
    threats while preserving Matter’s core principles of openness and
    interoperability.

\end{itemize}

\section{Preliminaries}\label{Prelims}

\subsection{Smart Homes}

Modern smart homes typically incorporate a diverse array of connected devices,
such as sensors, lighting systems, locks, and cameras, coordinated through
centralized hubs which function as control points that enable users to manage
and monitor devices on the home network. Historically, smart home networks were
fragmented as devices from different manufacturers often relied on proprietary
software and required dedicated controllers. As a result, it was common for
households to operate under multiple incompatible hubs that each managed a
subset of IoT devices within their respective networks. This reliance on closed
ecosystems not only complicated the management of smart homes, but also
restricted cross-device compatibility, hindering third-party development and
innovation.

The Matter protocol~\cite{matter}, developed by CSA, aims to address these
limitations by introducing a standardized, open framework for smart home
connectivity. A key feature of Matter is its ability to enable seamless
communication among devices from different manufacturers, regardless of their
underlying platforms. Operating at the application layer over IPv6, Matter is
agnostic to Layer 2 technologies and currently supports Ethernet and Wi-Fi for
high-bandwidth, always-on devices, as well as Thread~\cite{thread} for
resource-constrained, battery-powered devices. This design unifies communication
across diverse network topologies and device classes within a single IP-based
ecosystem.

\subsection{Matter Primer}

Development of the standard began in 2019 as Project Connected Home over IP
(CHIP), which CSA renamed Matter in 2021; Matter 1.0 and its certification
program launched in October 2022~\cite{matter-history,matter-launch}. The
ecosystem is backed by major platform providers, including Amazon, Apple,
Google, and Samsung SmartThings, as well as device and semiconductor vendors
such as IKEA, Signify, NXP, and Silicon Labs~\cite{matter-history}. Their support
has brought Matter controllers and end devices to consumer platforms and
commercial product categories ranging from lighting, smart plugs, and locks to
sensors, HVAC equipment, media devices, and bridges~\cite{matter-launch}. By
November 2024, CSA reported that Matter products had been deployed in millions
of homes worldwide~\cite{matter-adoption}, demonstrating the standard's
transition from an industry specification to an actively used ecosystem.

To make this happen, Matter defines a data model governed by a standardized interaction protocol.
Each participant is represented as a node, which authenticates peers
cryptographically and exposes functionality through endpoints. Each endpoint
contains clusters that define specific capabilities (\eg~on/off control,
temperature sensing). Nodes interact via encrypted messages to read attributes,
invoke commands, and subscribe to events. Following the Matter
specification~\cite{matter}, we use the term \textit{device} to refer to an IoT
gadget, and \textit{controller} to refer to the hub.

\begin{figure}[t]
    \centering
    \includegraphics[width=\columnwidth]{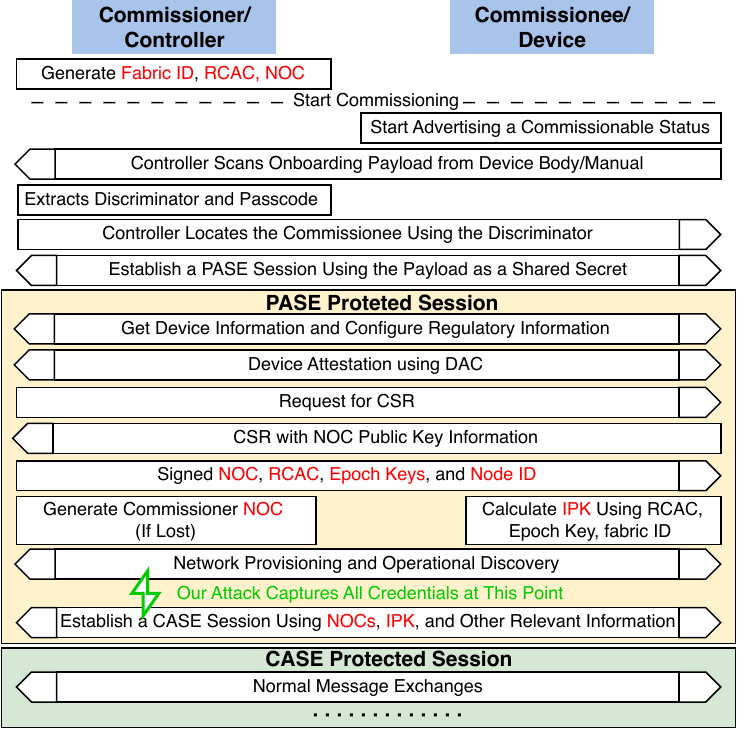}
    \caption{Device Commissioning (credentials in red).}
    \label{fig:commissioning}
\end{figure}

\subsubsection{Commissioning Protocol}

Matter devices communicate within a logical network called a fabric, which is
managed by a controller (\eg~a smart hub) that is responsible for onboarding new
devices, configuring them, and enforcing access control. To join a fabric, a
device undergoes commissioning, which establishes its permanent credentials and
a secure session with the controller. A simplified workflow of device enrollment
is shown in~\autoref{fig:commissioning}; we outline the main steps below,
omitting minor details for clarity.

The controller serves as the commissioner, while
the device being added to the fabric serves as the commissionee.
During commissioner initialization, a 64-bit \textit{Fabric ID} is randomly selected
to uniquely identify the fabric being created, 

along with a self-signed X.509v3 \textit{Root Certificate Authority (RCA)} certificate serving as the fabric's trust anchor.
Using this RCA, the commissioner signs a certificate for itself, named the \textit{Node Operational Certificate (NOC)}.
Note that every node in the fabric, including the commissionee, will have an NOC signed by RCA.

The commissioning process begins with the commissioner obtaining the onboarding payload from the commissionee---typically a QR code or alphanumeric string printed on the device or its manual. The payload encodes two essential values: (1) a discriminator, used to identify the commissionee during device discovery, and (2) a passcode, which proves ownership during authentication.

The commissioner and commissionee then run a \textit{Passcode Authenticated Session Establishment (PASE)} protocol, which leverages SPAKE2+~\cite{spake} to derive a shared session key without exposing the passcode in plaintext, establishing a secure channel for the rest of the commissioning process.

With the PAKE session established, the commissioner retrieves device and regulatory information (omitted for brevity) and then performs attestation to verify that the device is a certified Matter node. Each device holds a unique \textit{Device Attestation Certificate (DAC)}, signed by a manufacturer-controlled authority listed in the \textit{Distributed Compliance Ledger (DCL)}. The commissioner validates the DAC by issuing a challenge that the device signs with its private key. Notably, attestation is required only for the commissionee, not the commissioner; thus, any entity can act as a Matter controller, a property we exploit in this work.

After attestation, the device generates a \textit{Certificate Signing Request (CSR)} using its operational key pair, which the commissioner signs with its \textit{RCA Certificate (RCAC)} to produce the device’s NOC. The device also receives the RCAC, a 64-bit Node ID, and up to three Epoch keys. It then derives an \textit{Identity Protection Key (IPK)} from the second newest Epoch key and the compressed fabric ID; the IPK serves as a shared group key used in final session key establishment.

The protocol then performs network provisioning and operational discovery before initiating \textit{Certificate Authenticated Session Establishment (CASE)}. CASE, based on the Sigma protocol~\cite{sigma}, provides mutual authentication by exchanging the NOCs of the commissioner and commissionee, with the IPK protecting identities during the handshake and contributing entropy to session key derivation.

\subsubsection{Data Encoding}

Matter uses a Tag-Length-Value (TLV) format to encode both data at rest
(\eg~certificates) and data in motion (\eg~message exchanges). Similar to ASN.1
DER but simpler and more compact, Matter TLV represents typed and nested data
via a control octet defining the element’s structure. It supports canonical
ordering for consistent cryptographic operations and includes both primitive and
container types, enabling efficient representation of complex data.

\label{sec:cryptokeys}

Matter employs Elliptic Curve Cryptography (ECC) on the NIST P-256 curve, with
64-byte public keys and 32-byte private keys. Certificates follow the X.509
structure but are encoded in TLV rather than DER. Epoch keys, shared across the
fabric and TLV-encoded, contain a 16-byte random value and a timestamp, which
are essential for IPK recovery in our attack.

\subsubsection{Access Control}

Matter regulates access through fabric-scoped policies applied to device clusters---logical groupings of functionality. Each node maintains an Access Control List (ACL) composed of entries defining permissions and scope. Four privilege levels are defined: View (read-only), Operate (commands/actions), Manage (configuration), and Administer (full control, including ACL modification).

\subsection{Temporarily Shared Space}

Smart home adoption has grown rapidly, with the market projected to quadruple by the end of the decade~\cite{sh-market}. 
This trend extends beyond private residences: IoT technologies have seen similarly swift 
uptake in the hospitality sector~\cite{amazon-iot, august-iot, sm-gadgets-iot, practices-iot, benefits-iot, airbnb-hosts},
driven in part by the rise of short-term rental platforms like Airbnb, 
where guests increasingly expect the same smart amenities they enjoy at home~\cite{benefits-iot,practices-iot}.

The rapid proliferation of smart homes, combined with increasingly common shared
living arrangements, introduces serious and often overlooked security risks.
Unauthorized use and abuse of smart devices are already
widespread~\cite{smarthome-misuse}, with adversaries, including controlling
partners, family members, and short-term guests, actively exploiting IoT
ecosystems for surveillance and manipulation~\cite{partner-abuse}. These
real-world scenarios directly enable the threat we study: they routinely grant
malicious actors temporary, unsupervised physical access to a smart home,
creating a practical and highly realistic pathway for the attacks we
demonstrate.

\subsection{Memory Attacks in Unattended Environment}
\label{sec:physicalattack}

We survey several attack scenarios targeting the memory systems of embedded devices  
where attackers have physical access to the device, as satisfied in home-sharing environments.
These works deal with a range of memory types with different levels of protection.

It was commonly assumed that data is lost from dynamic RAM (DRAM) on power loss. 
However, due to the remanence property of DRAM, data can persist for a short time after power is removed. 
Halderman et al. showed that cold boot attacks can exploit this property to allow attackers with physical access 
to a device to recover sensitive data directly from memory~\cite{coldboot}.  
Their work showed that at room temperature, complete data loss is not immediate and by reducing the temperature of 
chips to -50°C (dry ice, which is readily available at -78°C), 99.9\% of data is retained for at least 1 minute. 
This work was expanded by Müller~\etal~\cite{frost}
to target mobile and embedded devices.

As a countermeasure to cold boot attacks, many secure systems now confine sensitive data to on-chip memory made of static RAM (SRAM), 
where data retention is minimal after a brief power loss---even at very low temperatures~\cite{targetedenc,zhang2016case}. 

However, researchers demonstrated that by supplying a steady voltage to the power pin of a System-on-chip (SoC), 
it is possible to preserve the state of SRAM across a power cycle~\cite{voltboot}. 
This technique, called Voltboot, effectively replicates cold boot behavior for SRAM, enabling full recovery of 
sensitive data from SRAM such as CPU cache and internal RAM.

Another attack, known as UntrustZone~\cite{untrustzone}, demonstrates that credential information---even when 
afforded high-level protection mechanisms such as storage in SRAM and safeguarding 
by Trusted Execution Environments (TEEs) like ARM TrustZone---remains vulnerable when an adversary gains physical access to the device. 
The researchers observed that SRAM undergoes gradual analog-level changes during routine software access. 
However, through physical intervention, these changes can be significantly accelerated. 
Consequently, secrets stored in SRAM become imprinted in the memory’s analog domain quickly, 
allowing attackers to recover them by analyzing the SRAM’s power-on state. 
The study successfully extracted TrustZone-protected AES keys from 13 different devices, 
all under 24 hours, well within the typical duration of a guest’s stay in a shared home environment.

While attacks on volatile memory must contend with its inherent impermanence, 
persistent memory, such as flash, can be equally impactful and significantly easier to exploit in unattended environments~\cite{cui2013firmware}. 
Flash retains its data over power cycles and often stores sensitive data like firmware and encryption keys, 
all of which remain even after shutdown. 
Attackers can exploit this persistence through a wide range of well-documented techniques 
to extract or modify flash contents such as connecting to exposed debugging ports like JTAG or 
using a Serial Peripheral Interface (SPI) flash readers such as Bus Pirate~\cite{recessim}.
Tools like Bus Pirate also support live I2C/SPI traffic capture, 
allowing passive observation and targeted analysis of sensitive data as it is accessed during normal operations.

Beyond these memory-specific attacks, chip-level vulnerabilities in
microcontrollers enable the extraction of protected secrets and compromise of
entire systems. In recent years, these weaknesses have been exploited through
physical techniques such as fault injection, power glitching, or voltage
manipulation~\cite{rp-challenge, esp32fi, walletattack, airtag}. For example,
the Raspberry Pi RP2350---equipped with secure boot and explicitly designed
to resist fault injection---was compromised during a hacking challenge
employing different fault injection methods~\cite{rp-challenge} which steal
the embedded secrets. Similarly, the ESP32-C3 and ESP32-C6, despite their advertised
resistance, were shown to allow full extraction of flash contents~\cite
{esp32fi}. Power-glitching attacks on the STM32F205 chip in the Trezor One
hardware wallet enabled full recovery of stored cryptocurrency~\cite
{walletattack}, while voltage glitching on Nordic’s nRF52832 allowed cloning
and modification of Apple’s AirTag~\cite{airtag}.

\section{Attack Overview}\label{sec:overview}

\subsection{Attack Model and Prerequisites}

We assume a motivated and technically experienced adversary who has been granted
legitimate, temporary physical access to a smart home. The attacker has a
specific interest in compromising a particular household and sufficient time to
carry out non-trivial attacks. For example, a controlling partner or family
member engaging in coercive surveillance~\cite{smarthome-misuse, partner-abuse},
a landlord monitoring tenants without consent, a visitor conducting pre-burglary
reconnaissance (\eg~under the guise of a repair visit), or an Airbnb guest
exploiting unsupervised access during their stay~\cite{spacelord}. Common to all
these scenarios is a period of unsupervised physical access motivated by a
specific interest in the household.

During this access period, we assume that the attacker exploits the unattended environment and 
is capable of extracting secrets stored on devices or controllers, 
an increasingly realistic threat, as discussed in~\autoref{sec:physicalattack}. 
For example, the attacker can employ techniques such as cold-boot~\cite{coldboot}
and fault injection~\cite{reviewFI} to exfiltrate
secrets from both DRAM- and SRAM-based memory, even when that memory is well protected in TEEs and other protected memory regions~\cite{untrustzone, riscv-FI, voltpillager, voltjockey1}.
Flash sniffing remains another viable attack vector, where an adversary can use a logic analyzer to intercept traffic between a microcontroller and its external XOR Flash memory~\cite{cui2013firmware,recessim}.

Additionally, we assume that a tiny covert device added to the fabric as a result of the attack can remain physically hidden within or near the victim’s home. 
For instance, the attacker might conceal it under a patio, inside a utility closet, or in any location 
unlikely to be inspected. 
This allows for persistent remote access to the smart home environment, even after the attacker’s physical presence has ended.

\subsection{Attack Goals}
\label{sec:goals}

The attackers' goal is to control the entire Matter fabric in a stealthy and
persistent manner, without requiring any product-specific reverse engineering
(\eg~firmware analysis). To this end, the attacker targets the open
characteristics of the Matter protocol. A covert device is embedded into the
fabric, enabling it to impersonate legitimate fabric entities while allowing the
victim to continue interacting with their devices without noticeable disruption.
Once deployed, the covert device allows the attacker to remotely monitor
controller--device traffic, inject falsified sensor data, and trigger any
Matter-permitted actions, even after leaving the premises and even for devices
added to the fabric at a later time.

\paragraph{Alternatives.} One might wonder why an attacker would not simply
 deploy a single covert out-of-fabric device, such as a hidden surveillance
 camera. We argue that such an approach is intrinsically limited: an
 out-of-fabric device can observe only what its own sensors capture. To obtain
 broader coverage, the attacker would need to deploy multiple out-of-fabric
 devices across different monitoring locations, each equipped with distinct
 sensing modalities---cameras, microphones, motion sensors, contact sensors, and
 others---significantly increasing both the cost and the likelihood of
 detection.

In contrast, the attack we aim to realize leverages a single covert in-fabric
device that automatically harvests information from all existing and future
Matter devices. Because the covert device impersonates legitimate fabric
members, it can ``sense'' the victim’s environment through data streams that
users expect their own devices to generate, even if they are added after the
compromise.

\begin{figure}[t]
    \centering
    \includegraphics[width=\columnwidth]{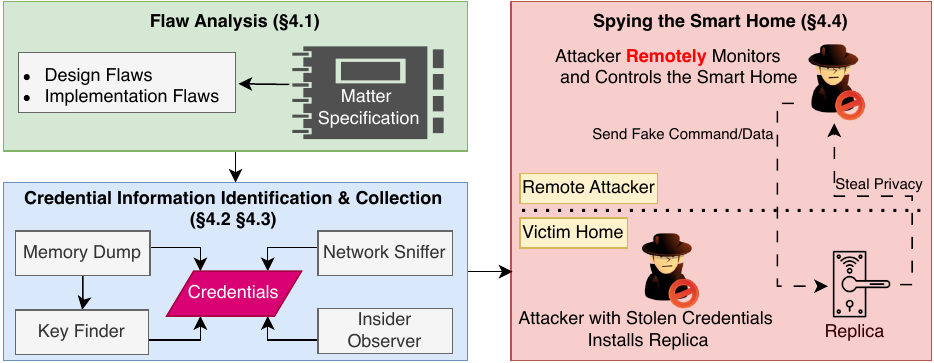}
    \caption{Attack Overview.}

    \label{fig:workflow}
\end{figure}

\subsection{Attack Workflow}
\label{sec:workflow}

Our work comprises five main components, as illustrated in~\autoref{fig:workflow}. 
First, we examine the Matter specification and uncover several design flaws and potential implementation weaknesses, 
which together establish the preconditions for our attack (\autoref{sec:flaws}). 
Second, we identify the critical credentials required to duplicate controllers and devices, 
and assess the feasibility of extracting such information (\autoref{sec:creidentify}). 
Third, we present the tools we developed to obtain these credentials, 
including a key finder and a network eavesdropping tool (\autoref{sec:credcollection}). 
Fourth, with the necessary credentials in hand, we demonstrate how to configure and clone a 
controller or device within the target environment,
and showcase a spectrum of attacks that can be executed using the replicated controller and device, 
including remote surveillance and data manipulation (\autoref{sec:spying}).

\section{Design}\label{sec:design}

\subsection{Flaw Analysis}
\label{sec:flaws}

Our attack depends on design (\textbf{D\#}) and implementation (\textbf{I\#})
flaws we identified through documentation and code reviews.

\paragraph{D1: No Commissioner Attestation.} 
While devices are required to undergo an attestation procedure to verify their Matter certification, 
controllers are not subject to an equivalent form of authentication and are implicitly trusted as fabric administrators. 
As a result, any entity can manufacture a Matter controller that talks to certified Matter devices.

\paragraph{D2: Credential Exposure.}
The Matter standard involves many group credentials that are shared across the fabric.
Once a device controlled by the attacker is enrolled in the system,
these group credentials can be easily harvested by the attacker,
including the Epoch keys, fabric ID, RCAC, and IPK.
Note that although Epoch keys need to be updated and rotated periodically,
these operations are fully synchronized among all the nodes.

In addition, network traffic may expose some of these credentials in plaintext.
For example, each device must broadcast its Compressed Fabric ID and Node ID using multicast DNS (mDNS).
This is necessary because a device can belong to multiple fabrics and 
it needs to advertise its presence to all its controllers.

\paragraph{D3: Indistinguishability of Nodes with Identical Credentials.}
During dynamic operational discovery, 
a responder advertises its compressed fabric ID, Node ID, and current IPv6 addresses via mDNS broadcasting after receiving a request.
Upon detecting a valid mDNS message, the initiator attempts communication using one of the advertised addresses.
As a result, if a node’s credentials are stolen, it can be easily impersonated within the fabric. 
An adversarial replica sharing the same Node ID but operating from a different IP address 
becomes indistinguishable from the legitimate node.

\paragraph{I1: Plain-text Credential Exposure.}
Existing Commercial Off-the-Shelf (COTS) IoT devices generally lack both hardware- and software-level protections, 
rendering them susceptible to memory attacks that extract sensitive information from flash memory, DRAM, SRAM, 
and CPU caches~\cite{coldboot,voltboot,targetedenc,zhang2016case,recessim,cui2013firmware,untrustzone}. 

It is not uncommon for cryptographic keys to remain in memory after use, 
thereby unnecessarily prolonging their exposure to potential attackers.

\paragraph{I2: Commissioning Non-Certified Devices.}
To foster a trusted environment, 
Matter mandates that only certified devices be commissioned into a fabric. 

However, most COTS controllers such as Homey Pro and Google Nest provide the option to override the validation result, 
allowing the commissioning of devices developed by non-certified parties.

\paragraph{I3: Long-Term Credentials.}
Node credentials are not only easily accessible, 
but are also commonly configured as long-term secrets, 
thereby amplifying the impact of their compromise. 
For instance, in our experiments, the default expiration period for an NOC exceeded 10 years.

\paragraph{I4: Insecure Local Interface Exposure.}
Commercial products frequently expose local interfaces, such as USB ports, for
maintenance purposes. This allows attackers with physical access to extract the
firmware or filesystem, potentially recovering sensitive credentials.

\subsection{Credential Information Identification}
\label{sec:creidentify}

To clone a Matter controller or device, one can capture all relevant credentials 
at the conclusion of the CASE protocol---particularly the session key, which is used to encrypt the messages. 
However, we do not choose this approach for several reasons. 
First, session keys are short-lived credentials as they can be actively terminated,
for example, due to limited resources,

reducing their usefulness for long-term attacks. 
Second, a device may maintain multiple session keys concurrently, 
requiring non-trivial reverse engineering efforts to accurately map each key to its specific usage context. 
Third, and perhaps most practically, a session key is merely a 32-byte random value 
with no associated metadata or structure. 
As such, identifying the correct key in memory is particularly challenging, 
as there is no signature or distinguishing characteristic to aid in its detection.

To avoid these issues, we instead choose to capture credentials necessary for
the execution of the CASE protocol and reinitiate CASE as needed (see the green
label in~\autoref{fig:commissioning}). These credentials exhibit more persistent
and identifiable properties without sacrificing utility: (1) they are
long-lived; (2) they are sufficient to start a new CASE protocol to negotiate a
fresh session key; and (3) they follow a specific format or can be derived from
other recognizable credentials. In~\autoref{tab:creds}, we list all these
credentials including how they can be captured or derived. The concrete tools to
capture them are discussed in~\autoref{sec:credcollection}.

\begin{table*}[h]
\scriptsize
\caption{Credential Information and How to Obtain it (E: Extracted; D: Derived).}
\label{tab:creds}
\centering
\begin{adjustbox}{width=\textwidth}
\begin{tabular}{l|cll}
\hline
\textbf{Credential Name} & \textbf{Method} & \textbf{Extraction Sources} & \textbf{Derivation Sources} \\ \hline \hline
Fabric ID                & E                          & NOC; Insider Observer (\autoref{sec:partiallytrusted})              & -- \\ \hline
Compressed Fabric ID     & E/D                        & Network Sniffer (\autoref{sec:networksniffer})                      & Fabric ID and RCAC Public Key \\ \hline
RCAC                     & E                          & Memory Dump (\autoref{sec:memdump})/Key Finder (\autoref{sec:keyfinder})/Insider Observer (\autoref{sec:partiallytrusted})                           & -- \\ \hline
RCAC Private Key         & E                          & Memory Dump (\autoref{sec:memdump})/Key Finder (\autoref{sec:keyfinder})                                            & -- \\ \hline
Epoch Key Set            & E                          & Memory Dump (\autoref{sec:memdump})/Key Finder (\autoref{sec:keyfinder})/Insider Observer (\autoref{sec:partiallytrusted})                           & -- \\ \hline
IPK                      & D                          & --                                                                & Epoch Key Set and Compressed Fabric ID \\ \hline
NOC                      & E                          & Memory Dump (\autoref{sec:memdump})/Key Finder (\autoref{sec:keyfinder})                                            & -- \\ \hline
NOC Private Key          & E                          & Memory Dump (\autoref{sec:memdump})/Key Finder (\autoref{sec:keyfinder})                                            & -- \\ \hline
Node IDs                 & E                          & Network Sniffer (\autoref{sec:networksniffer})                      & -- \\ \hline
\end{tabular}
\end{adjustbox}
\end{table*}

\begin{itemize}[leftmargin=*]

\item \textit{Fabric ID}: It is essential for constructing the Destination Identifier 
in the initial Sigma1 message of the CASE session establishment.
It can be obtained through several means: 
by extracting it from an NOC, and by retrieving it from an attacker-controlled device that has been added to the target fabric.

\item \textit{Compressed Fabric ID}: This is a 64-bit obfuscated representation
of the fabric ID. It is used along with an epoch key in the derivation process of the IPK.
It can be captured by sniffing operational discovery packets or calculated using the Fabric ID and the RCAC public key.

\item \textit{RCAC}: 
It is required to construct the Destination Identifier, a critical part of the Sigma1 message during CASE session establishment.
It can be extracted from an attacker-enrolled device in the same target fabric,
or retrieved from the memory or flash storage of an existing fabric member.

\item \textit{RCAC Private Key:} It is used to issue the NOCs of
 devices being commissioned to the fabric. It can be captured from the memory
 or flash of the fabric's commissioner.

\item \textit{Epoch Key set}: This key set is used to derive the IPK.
Similar to RCAC, it can be captured on an attacker-enrolled device in the target fabric
since both of them are group credentials.

\item \textit{IPK}: It is essential for constructing the Destination Identifier 
in the initial Sigma1 message of the CASE session establishment. 
Additionally, the IPK serves as key material in the session key derivation during CASE.
It can be derived from the Epoch Key and Compressed Fabric ID.

\item \textit{NOC}: 
It is required for authentication within the fabric during the CASE session establishment, 
specifically in the Sigma2 and Sigma3 messages.
It can be captured from the memory or flash of a fabric member.

\item \textit{NOC Private Key}: 
It is required for authentication within the fabric during the CASE session establishment.

It can be captured from the memory or flash of a fabric member.

\item \textit{Node IDs}: A Node ID is essential for constructing the 
Destination Identifier in the initial Sigma1 message during CASE session establishment, enabling communication with the corresponding device.
They can be captured by sniffing operational discovery packets.

\end{itemize}

\subsection{Credential Information Collection}
\label{sec:credcollection}

This section presents the techniques 
for locating and retrieving the aforementioned credential information.

We focus on credentials directly extractable from the device and omit algorithms to recover derivable ones
since the same algorithms are publicly documented by CSA~\cite{matter}.
For instance, the IPK can be recovered using the Epoch Key and the Compressed Fabric Identifier.

\subsubsection{Memory Dump}
\label{sec:memdump}

An attacker with physical access to the Matter controller in an unsupervised environment
can perform memory extraction attacks such as cold boot, or
access flash contents by interfacing with exposed hardware using flash readers or sniffers~\cite{recessim,cui2013firmware}. 
The hypothesis is that credentials, even encrypted at rest,
have to be decrypted into the RAM for run-time computation.
Therefore, we use existing well-established methods to capture both volatile and non-volatile memory.

The resulting memory dump or sniffed traffic is then analyzed using the Key Finder described in~\autoref{sec:keyfinder} 
which extracts the needed credentials.
Alternatively, an attacker can also directly access the local interface of the device to retrieve the firmware or filesystem, which may contain credentials in plaintext.

\subsubsection{Key Finder}
\label{sec:keyfinder}

This module can extract RCAC, NOC, and their private keys from memory dumps or flash transactions.
Certificates such as the RCAC and NOC are X.509v3-compliant and 
encoded using Matter's TLV format. 
Our Key Finder locates candidate certificates by scanning for headers and footers corresponding to both DER and TLV encodings, 
constrained to expected length ranges. 
Each candidate is validated, and relevant components (\eg~the public key, a 64-byte value prefixed by 0x04) 
are extracted using certificate encoding/decoding tools provided by Matter's reference implementation~\cite{chiptool}. 
This ensures precise and reliable identification of certificates.
We check the subject and issuer fields to differentiate NOC and RCAC.
Specifically, the RCAC is self-signed with an identical subject and issuer,
while the NOC is issued by the RCAC.

Once a valid NOC/RCAC public key is identified, 
it is used to locate its corresponding private key in memory. 
The private key is often stored in raw form (\ie~not encoded), making it harder to identify. 
According to the Matter Specification, the relevant key pairs use the NIST P-256 elliptic curve. 
Candidate private keys are therefore identified as 32-byte, word-aligned, high-entropy values 
within the valid range for the P-256 curve. 
A candidate private key is confirmed by reversing the elliptic curve operation conducted by its associated public key.

\subsubsection{Network Sniffer}
\label{sec:networksniffer}

Commissioned devices in a Matter fabric advertise their presence and availability 
by sending plaintext mDNS advertisements containing two key pieces of information: 
(1) Compressed Fabric ID, and (2) Node ID.

Controllers use this information to discover the device’s current IPv6 address and initiate communication.
Any attacker device connected to the same Wi-Fi network as the fabric can passively capture 
these mDNS advertisement packets via service-discovery tools such as Bonjour~\cite{bonjour} or Avahi~\cite{avahi}, 
allowing them to identify all Node IDs present within the fabric.

\subsubsection{Insider Observer}
\label{sec:partiallytrusted}

In a shared smart home setting, a guest may be granted permission to enroll their own Matter devices. 
Once an attacker-controller device joins the fabric, it can easily steal group credentials belonging to the fabric.

The credentials recoverable through this method include the Fabric ID, RCAC, and the Epoch Key Set.

\subsection{Attacking the Smart Home}
\label{sec:spying}
\label{sec:devduplication}

When sufficient credentials have been collected, an attacker can build
duplicates of both controllers and devices. Specifically, he can import these
credentials into any Matter-compliant hardware. Once the credentials are
embedded, the attacker launches the controller/device replica, verifies that
it functions correctly, and ensures that the legitimate controller UI does
not show any anomalies (\eg~displaying an unknown device). Such anomalies are
unlikely due to flaw D3 as discussed in~\autoref{sec:flaws}. Next, the
attacker implants a backdoor into the clones through which they can remotely
control the victim controller/device later. Finally, these backdoored clones
are discreetly deployed within the target premises. These clones now operate
as legitimate fabric members, enabling unauthorized access, persistent
surveillance, and active manipulation within the Matter network, all under
the control of an attacker who has already vacated the environment.

\paragraph{Controller Replica.} 
From the perspective of fabric members, the legitimate controller and its replica are indistinguishable. 
The replicated controller can interact with all members of the fabric under the same level of trust and privileges  as the original controller. 
This includes both active and passive information retrieval (\eg~via subscribe/publish mechanisms), 
issuing operational commands (\eg~turning devices on/off), 
and performing administrative tasks (\eg~modifying ACLs). 
Notably, the attack can persist even after the attacker leaves, 
as newly added devices continue to trust the replica---given that their 
NOCs are signed by the same trusted CA.

The controller replica remains stealthy within the fabric. 
Although a legitimate controller may query any device to obtain a list of all controllers 
that have access to it. 
However, because the replica uses the same credentials as the legitimate controller, 
it is treated as the same entity and not recognized as a distinct controller instance.

\paragraph{Device Replica.} 

Before leveraging the stolen credentials to establish new CASE-authenticated sessions, 
the replicated device must be configured to communicate with the designated controller. 
This is achieved by explicitly programming an all-allow ACL. 
Once done, the replica, impersonating the legitimate device, 
can carry out unauthorized operations such as injecting falsified data into the fabric. 
Because the legitimate and replicated devices share identical credentials, 
other fabric members cannot distinguish between them, thereby enabling stealthy and persistent operation.

Even though they share the same credential, 
they are physically distinct hardware entities with unique IP addresses. 
According to the protocol, each device periodically broadcasts an mDNS record to 
advertise its presence and discovery information to peers. 
These records include three fields: \texttt{<Hostname>}, a Matter service
instance identifier of the form \texttt{<CompressedFabricID-NodeID>} with
the \texttt{.\_matter.\_tcp} suffix, and \texttt{<IP>}.
Here, the hostname is encoded as \texttt{<MAC>.local},
the second field represents the device’s Matter identity, 
and the IP field lists the IPv6 addresses assigned to the device. 
Consequently, these records differ in hostname and IP address.
From the controller’s perspective, 
these broadcasts are treated as originating from the same device. 
This ambiguity introduces uncertainty in communication, 
as the controller may arbitrarily select either the victim device or its replica when establishing a session.
To overcome this limitation and ensure precise control, 
we introduce a Man-in-the-Middle (MiTM) box. 
This intermediary regulates network connectivity and enables
consistent interception or redirection of traffic, 
thereby reliably supporting the full spectrum of attacks described in this work.

\begin{figure*}[t]
\centering
\includegraphics[width=\textwidth]{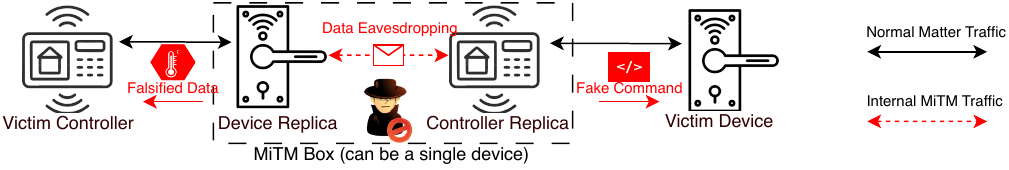}
\caption{Building a Man-in-the-Middle Box with Replicas.}
\label{fig:mitm}
\end{figure*}

\paragraph{MiTM Box.} 
Our MiTM box is a single physical device that runs multiple Matter instances including a replica controller and a set of replica devices, 
each corresponding to a specific victim device. 
As illustrated in~\autoref{fig:mitm}, these replicas are specifically positioned 
to regulate and intercept all communication.

When the MiTM box is in effect, the victim controller establishes a session 
with the replica device, while the replica controller maintains a separate session with the corresponding victim device. 
Communication between the victim endpoints is transparently relayed through the MiTM infrastructure.
While the replica controller and replica devices are logically separate,
they can be implemented on a single piece of hardware.

As previously discussed, the victim controller may unpredictably select
either the legitimate device or its replica when initiating communication. To
enforce the communication topology shown in~\autoref{fig:mitm}, additional
mechanisms are required. The replica controller, being fully controlled by the
attacker, is configured to communicate exclusively with the legitimate device.
However, the victim controller may cache the IPv6 address of devices to avoid
repeated mDNS queries, resorting to discovery only when the cached address
becomes unreachable (\eg~after an address change). Since the controller has
previously communicated with the legitimate device, it retains its IPv6 address
in the cache. To intercept this communication, the replica must take this IPv6
address. We achieve this via a Neighbor Discovery Protocol (NDP) poisoning
attack: the replica sends forged neighbor advertisements to the controller,
binding the legitimate device’s IPv6 address to the replica’s MAC address,
thereby redirecting subsequent traffic to the replica.

\paragraph{New Commissioning.} 
After departing from the shared home environment,
an adversary

retains the capability to replicate subsequently commissioned devices.
This is because the MiTM box can capture a new device’s Node ID via the network sniffer
and then sign a new NOC with RCAC private key for the replica,
thereby allowing the replica to impersonate the newly commissioned device.

\section{Implementation}
\label{sec:implementation}
\label{sec:attacksystem}

Our prototype implementation comprises a suite of tools to
(1) identify, extract, and derive credentials from real Matter-compliant devices;
(2) replicate Matter controllers and devices within victim smart home environments; and
(3) program the duplicated devices to facilitate various attacks, 
including the construction of a MiTM box and implementation of malicious logic.

\begin{figure}[t]
    \centering
    \includegraphics[width=\columnwidth]{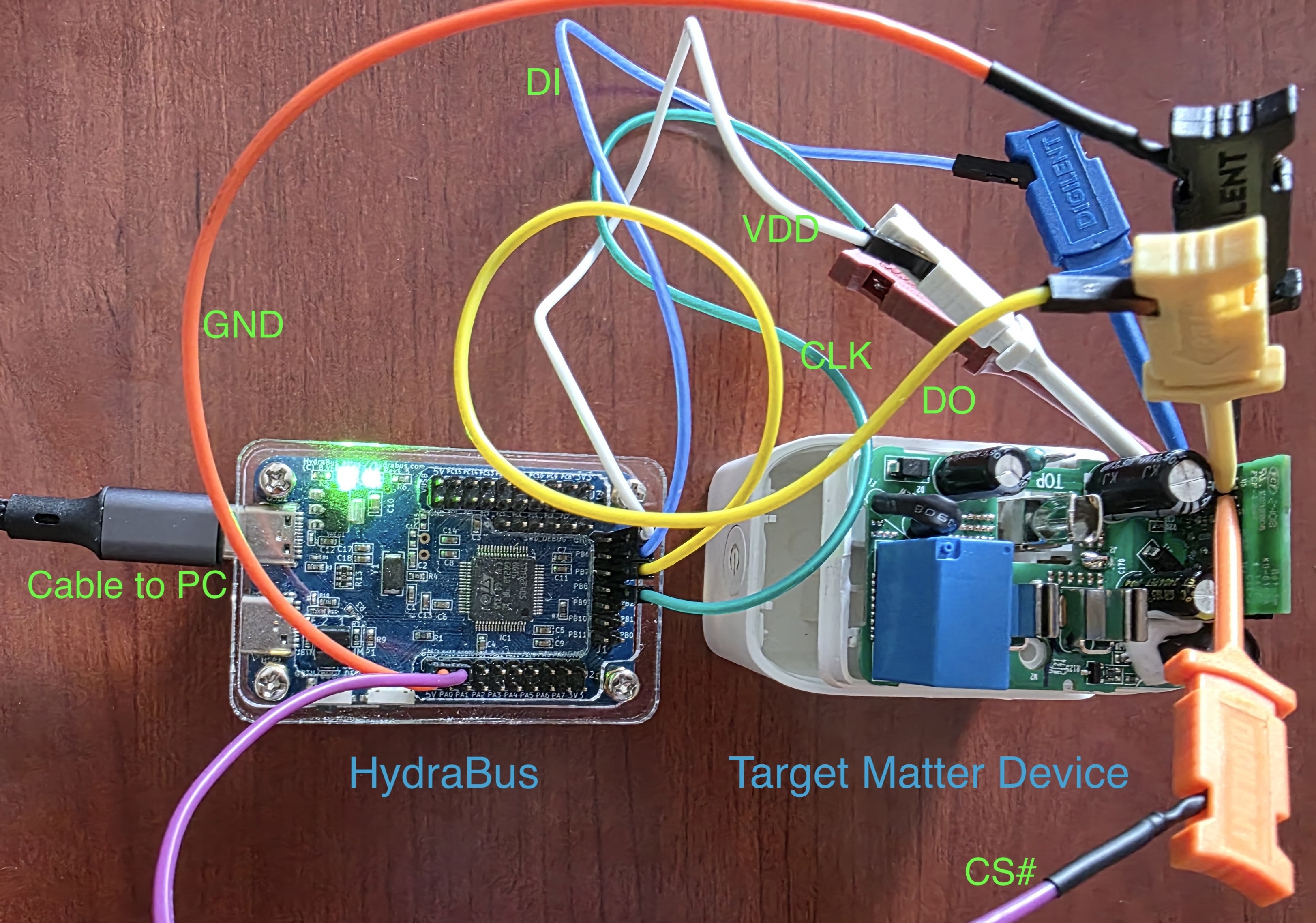}
    \caption{Flash Dump Setup.}
    \label{fig:flashdump}
\end{figure}

\paragraph{Memory Dump.} 
We implemented memory dumping tools for both RAM and flash storage. 
For RAM, we did not reproduce cold boot or VoltBoot attacks directly, 
as these attack mechanisms are orthogonal to the focus of our work. 
Instead, we simulated physical memory extraction by 
deploying the LiME kernel module~\cite{lime} on the target device,
which actively acquires a complete snapshot of volatile memory on the target.
For flash, we used HydraBus~\cite{hydrabus} to dump flash offline.
Our prototype supports popular flash (NOR and NAND) and interface (SPI, QSPI, and I2C) combinations,
covering the majority of flash deployments in the embedded market~\cite{spi1, spi2, spi3}.

In~\autoref{fig:flashdump}, we show the hardware setup
to read out flash from a commercial Matter-enabled smart plug.

\paragraph{Network Sniffer.}
We developed a Python-based mDNS sniffer that
monitors the local network for mDNS service advertisements from Matter-compliant devices. 
It parses incoming discovery packets to extract operational node identifiers, 
enabling passive enumeration of devices.

\paragraph{Insider Observer.} This is an attacker-enrolled device whose goal is to
harvest group credentials such as Epoch Keys.
We adapted example firmware from CSA’s official Matter repository~\cite{chiptool}
for the Espressif ESP32-S3-DEVKITC-1-N8 board, instrumenting the firmware
to automatically extract credentials provisioned by the legitimate controller during commissioning.

\paragraph{Key Finder.} The Key Finder is implemented as a Python script that 
leverages Matter’s TLV parsing library to extract structured operational certificates from memory dumps.

It then scans for high-entropy byte sequences as candidates of private keys.
A key pair is confirmed when the private key can reverse the elliptic curve cryptographic operations
performed by the public key stored in the corresponding certificate.

This allows us to collect the fabric's RCAC, 
the target devices' NOCs and their associated private keys, 
the fabric's Epoch Keys, and the Node IDs of devices commissioned to the fabric.
For other derivable credentials, we follow the original algorithms defined in the Matter specification~\cite{matter}.
In particular, the Fabric ID is acquired from the NOCs; 
the Compressed Fabric ID is computed using the Fabric ID and the RCAC's public key; 
and the IPK is derived from the Epoch Keys and the Compressed Fabric ID.
In case a strongly protected device prevents an attacker from extracting its credentials,
we leverage the fabric’s RCAC private key to generate and sign a new NOC.

To manage these credentials, we developed a script that consolidates them
into a configuration file conforming to the structure and format defined in the Matter specification.

\paragraph{Controller/Device Duplication.}
Matter standardizes device capabilities through the use of clusters. 
To streamline the duplication process for both controllers and devices, 
we developed templates for controllers and ten popular device types, 
including smart light bulbs, door locks, thermostats, energy management devices, and others.
Attackers simply need to identify the type of victim device
and apply the corresponding template and credentials to simulate the real hardware.
This highlights how easy it is to duplicate product functionality in the Matter ecosystem.
Our prototype builds upon the CSA's official open-source controller implementation,
commonly known as chip-tool~\cite{chiptool}, along with other reference projects for devices. 
For each target, our toolchain automatically imports the 
credential file mentioned before into the corresponding project and 
compiles the firmware for the appropriate development board.

\paragraph{Replica Augmentation.}
In addition to simulating the original functionality,
we augment replicas with additional logic to facilitate attacks.
First, the controller replica is equipped with an mDNS sniffer that periodically scans the local network to retrieve an 
updated list of Node IDs for devices within the fabric. 
This enables the replica to detect and control newly added devices, even
after the attacker has long vacated the shared space.
It also grants all ACL permissions to device replicas, ensuring they remain operable.

Second, we implement mechanisms to maintain the MiTM topology shown
in~\autoref{fig:mitm}. Our MiTM box, comprising both the controller and device
replicas, identifies the IPv6 addresses of both the legitimate controller and
device. It then configures Bettercap~\cite{bettercap} to perform NDP spoofing by
setting the legitimate controller’s IPv6 address as the target and the legitimate device’s IPv6
address as the spoofed neighbor. Bettercap subsequently sends forged Neighbor
Advertisement messages to the legitimate controller, causing it to associate the replica device's
MAC address with the legitimate device’s IPv6 address. To successfully intercept the
traffic, and prevent it from reaching the legitimate device, IPv6 packet
forwarding is disabled on the MiTM box. After setup, the MiTM box maintains
separate sessions with the respective hardware peers. The replicas decode and
re-encode messages as they forward traffic between legitimate devices and
controllers.

Third, we inject malicious payloads into both controller and device replicas to
launch actual attacks and establish remote control capabilities through Command
and Control (C2) servers. While our prototype uses the built-in Wi-Fi interface
for remote communication, a more advanced adversary could install a cellular
module to evade local Wi-Fi network scanners, enhancing stealthiness. On the device
replica, the payload enables the attacker to manipulate the data transmitted to
the controller and other devices (\eg~falsifying sensor readings, injecting
fabricated status reports). On the controller replica, the payload supports the
following capabilities: (1) exfiltrate sensitive data returned by legitimate
devices; and (2) issue unauthorized operational commands to the smart home
(\eg~unlocking doors and disabling alarms).

\section{Experiments}\label{Experiment}

We evaluate the proposed attacks and supporting tools
within a simulated smart home environment, consisting of a
combination of real-world COTS and
self-hosted Matter controllers and devices.
We aim to answer the following research questions:
\textbf{RQ1-Feasibility:} Can attackers extract Matter credentials from
Matter-compliant hardware in a device-agnostic manner?
\textbf{RQ2-Impact:} Can the MiTM box
introduce serious security and privacy threats to the victim smart home?
\textbf{RQ3-Stealthiness:} Can the attack be carried out
without disrupting legitimate operations or triggering noticeable indicators of compromise?

\begin{figure}[t]
\centering
\includegraphics[width=\columnwidth]{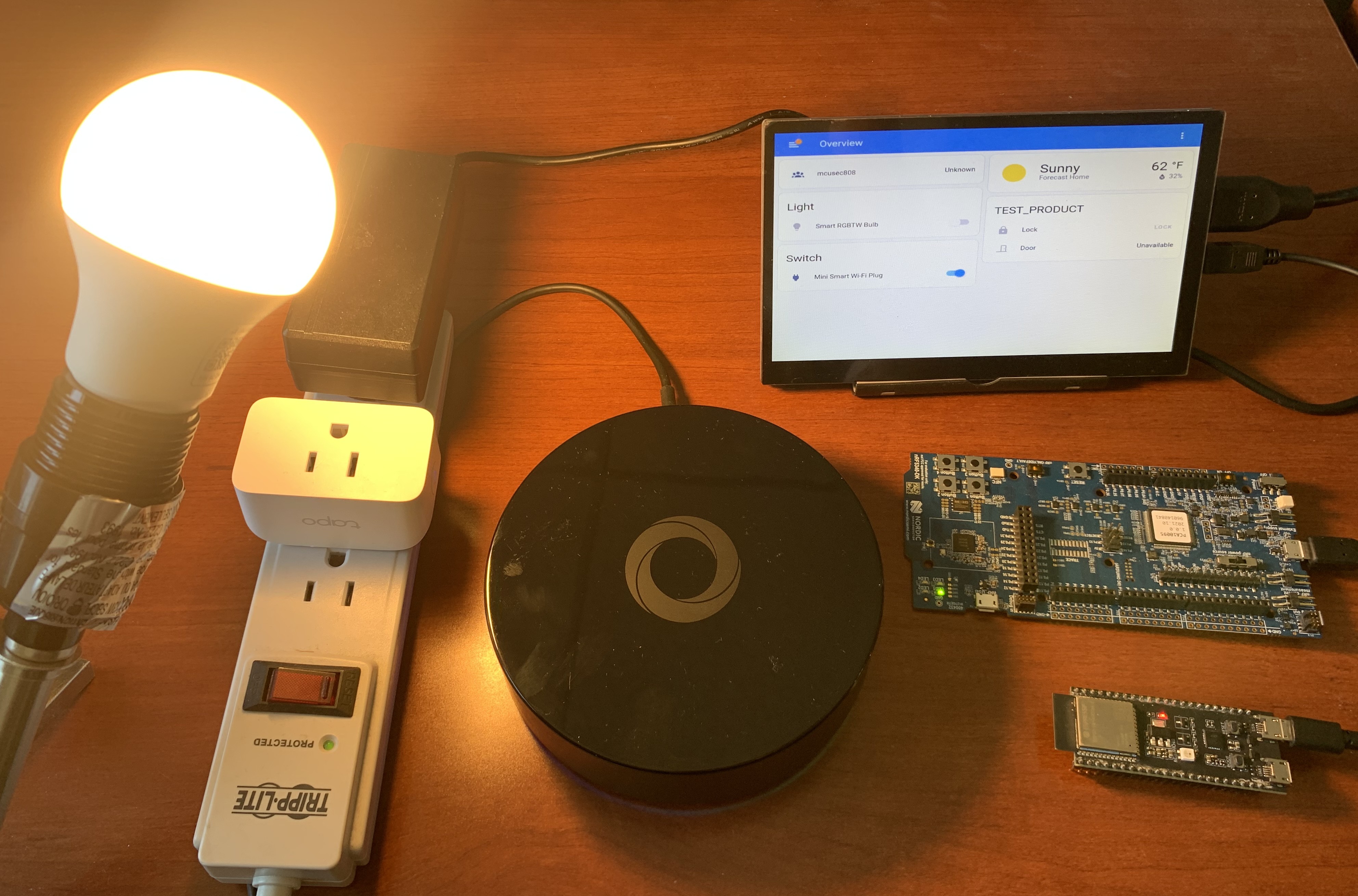}
\caption{Matter Environment Setup.}
\label{fig:env}
\end{figure}

\paragraph{Matter Environment Setup.}
\label{sec:targetsystem}
We deploy a controlled experimental setup to emulate a real-world Matter smart
home (\autoref{fig:env}). The setup includes two controllers: an open-source
Matter controller (Controller A) and a commercial closed-source Matter
controller (Controller B). The environment also includes a diverse set of
devices: two commercial Matter-enabled products (a smart plug and a smart bulb)
and Matter-compliant development boards (Espressif ESP32-S3 and Nordic nRF5340)
running open-source firmware for lighting, light-switch, and lock device types.
Including both an open-source controller and a consumer-oriented commercial
controller allows our evaluation to span different Matter deployment models.
Our MiTM box is implemented on a Raspberry Pi 4, which is a common platform for
self-hosted smart home controllers.

\subsection{RQ1: Credential Acquisition}

\subsubsection{Credentials from the Controller}

As a proof of concept, we deployed a LiME kernel module on Controller A.
Using LiME, we acquired a memory snapshot totaling 1.9~GB, consistent with the
RAM capacity of our Raspberry Pi device.
The acquisition process was completed in approximately 35 seconds.

From the controller's memory dump, our Key Finder identified five RCACs,
three NOCs, and one NOC private key.
Locating the certificates required only a few seconds,
and the exhaustive search for private keys completed in
approximately 12 hours---well within the typical duration of
a guest's stay in shared living environments.
By matching their issuer and subject fields,
we determined that the three NOCs were associated with a single RCAC.
These NOCs corresponded to two active devices under test and the controller itself.
The remaining four RCACs were not actively in use.
Upon further analysis, we discovered that these were retrieved from a device
that stores all trusted RCACs, including those generated during earlier experiments.

Through key matching, we located one private key corresponding to the controller's NOC.
As expected, no private key associated with any device was found, since
private keys should remain exclusively with their respective owners.
We did not find the RCAC private key either, since it should only
reside in memory during a short period of commissioning
to mitigate disclosure risks.
However, by inspecting the controller's flash memory contents,
we located the RCAC private key in a configuration file encoded using Matter TLV.

For Controller B, we discovered that an exposed USB interface provides access to
onboard storage. Using the platform's recovery interface, we obtained the
controller's persistent configuration data. The relevant configuration file
followed the same general key--value structure used by the Matter reference
implementation, allowing us to identify fabric-specific NOC and RCAC entries.
Notably, the RCAC key pair was also stored in plaintext within the configuration
data, consistent with the reference implementation.
Using this information, we successfully extracted the RCAC and its key pair,
along with the NOC, IPK, Fabric ID, and Node ID.

\subsubsection{Credentials from the Device}

For device-side extraction, we carefully opened the case of a commercial
Matter-enabled smart plug without causing any damage to its internal components.
The printed circuit board (PCB) exposed a 32-Mbit SPI NOR flash memory device.
The entire process of opening the device, inspecting the PCB, and
identifying the relevant components took less than ten minutes.
We connected six signals (\texttt{VDD}, \texttt{GND}, \texttt{DI}, \texttt{DO},
\texttt{CLK}, and \texttt{CS\#})
to HydraBus multi-tool hardware~\cite{hydrabus} (hooking HydraBus
to the flash chip required no more than five minutes),
which we used to extract flash memory, as shown in \autoref{fig:flashdump}.
Running at~42 MHz, we dumped the flash contents in 580~KB fragments
and then reassembled them into a single complete 4~MB image
in under five minutes. The extraction process took us approximately
20 minutes in total.

From the device's flash memory, our Key Finder extracted, in less than 5 minutes,
the complete certificate chain, comprising the RCAC, NOC,
and the corresponding NOC private key, as well as the Epoch Key Set used to
derive the Fabric's IPK.
Notably, the NOC private key was stored in plaintext, Base64-encoded, within a
configuration file, concatenated with the NOC public key following Matter's TLV
encoding format.
In addition to key materials,
we recovered the Fabric ID and Node ID from the subject field
of NOC certificates.

Similar results were observed on the two development boards running three
different firmware samples.

\subsection{RQ2: Smart Home Hacking}

\paragraph{Unauthorized Control and Surveillance.}
We installed the collected controller credentials, including the RCAC, NOC,
NOC private key, Fabric ID, and Fabric IPK,
into a chip-tool instance on Raspberry Pi 4, serving as the controller replica.
Acting as the communication initiator, we used the built-in chip-tool commands to
establish CASE-authenticated sessions with active devices.

As expected, we were able to selectively connect to any device in the fabric,
including two commercial products and two development boards running example
firmware.
Using these sessions, we successfully issued operational commands to the
devices and captured their responses.
Notably, this included devices commissioned after the controller was duplicated,
demonstrating that the controller replica retains access to
devices enrolled after the adversary has left.

\paragraph{Falsified Data Injection.}
We installed the collected device credentials from a commercial Matter-enabled
smart plug and three firmware examples
into the corresponding device templates, which served as device replicas.

By broadcasting the same Fabric ID and Node ID,
each replica successfully established a CASE-authenticated session with other
members of the Fabric,
replacing the functionality of the original device.

The replicas were able to inject falsified data into the controller and
intercept all commands issued to the original device.

As discussed in~\autoref{sec:spying}, both the legitimate device and its replica
multicast their respective IP addresses, creating ambiguity for the controller,
which may arbitrarily select one for communication.
This behavior was confirmed in our experiments: when both devices were powered
on simultaneously, the controller appeared to randomly select one.
Once selected, the controller maintained communication with that device
consistently, switching to the other only when the first instance went offline.
This behavior reduces the reliability of the attack and increases the
likelihood of detection.
For example, if the replica is selected and the legitimate device becomes
unresponsive, the user may notice the anomaly and initiate a security
investigation.

\paragraph{MiTM Attack.}
The MiTM box addresses the interference described above.
By performing an NDP poisoning attack on the legitimate controller, we
associated the replica device's MAC address with the legitimate device's IPv6
address, reliably enforcing the desired connection topology illustrated
in~\autoref{fig:mitm}. In our experiments, Bettercap helped the MiTM box
reliably establish the desired topology whenever the smart home network was
rebooted.

\subsection{RQ3: Stealthiness Evaluation}

After leaving the shared environment, the adversary can conceal the MiTM box
either within the premises or along its perimeter---for example, hidden beneath
a patio or inside a utility closet. Placing adversarial hardware along the
perimeter enhances stealth but imposes power constraints due to limited battery
capacity. In contrast, placing such devices indoors provides easier access to a
continuous power supply but increases the risk of detection by the homeowner.
The proposed attack requires only a single hidden MiTM box while still
harvesting traffic from all Matter devices in the fabric. This provides far
greater stealth than an approach that would require the adversary to deploy
separate sensing devices for every data type and location of interest.

In terms of system operation, we observed no noticeable service disruption or
anomalous behavior during either active attack phases or dormant phases
(\ie~passive MiTM relaying). To quantify the impact on normal operation, we
measured round-trip time (RTT) across 1,000 command--response exchanges under
three conditions: baseline (no MiTM), passive relay, and active message
modification. The baseline exhibited a mean RTT of 9.4 ms ($\sigma = 1.5$ ms),
while passive relay and active modification resulted in 18.4 ms ($\sigma = 2.2$
ms) and 19.6 ms ($\sigma = 2.3$ ms), respectively. The added latency, primarily
due to the additional Wi-Fi hop, remains below the threshold of human
perception. Furthermore, the 1.2 ms difference between passive relay and active
modification is negligible and falls within typical network jitter in smart home
environments.

From a network perspective, the replicas must remain connected to the home
network to maintain Matter functionality. This may raise suspicion if the
homeowner is familiar with the expected device layout and conducts a network
scan to identify unauthorized nodes. We elaborate on such detection strategies
in~\autoref{Prevention}. However, this concern is significantly reduced when the
home is occupied by a typical tenant who lacks detailed knowledge of the home
network topology.

\section{Discussions}\label{Prevention}

Matter brings meaningful innovation to the smart home ecosystem by 
enabling cross-vendor interoperability and embracing an open standard where security does not rely on obscurity. 
Its protocol specifications and reference implementations are fully transparent and publicly available, 
encouraging broad-based scrutiny.
This openness benefits all stakeholders: 

manufacturers can build products upon a well-understood and vetted foundation; 
end users are reassured by the availability of independent security verification rather
than reliance on proprietary secrecy;
ethical hackers are empowered to discover and address vulnerabilities.
Collectively, these innovations contribute to a more secure and trustworthy smart home environment.

Nevertheless, the same transparency that fosters collaboration and innovation 
also grants sophisticated adversaries the opportunity to study the system in depth. 
Armed with this knowledge, attackers can exploit Matter fabrics in a \emph{vendor-agnostic} manner
by taking advantage of the mandatory compliance requirements across devices. 

\subsection{Mitigation}
In~\autoref{sec:flaws}, we systematically identify a set of design-level (\textbf{D1–D3}) 
and imple-mentation-level (\textbf{I1–I4}) flaws. 
When combined, these issues enable powerful attacks with significant real-world consequences, as demonstrated in this work.
To mitigate these risks while preserving the advantages of Matter’s open architecture, 
we propose four countermeasures (\textbf{C\#}),
each designed to address a specific group of flaws uncovered in our analysis.
Among these, \textbf{C1} and \textbf{C3} can be readily integrated into the Matter specification, 
whereas \textbf{C2} and \textbf{C4} fall outside the scope of the specification and can be implemented through external mechanisms.

\paragraph{C1 $\rightarrow$ D3: Binding Credentials to Physical Features.} 
A fundamental cause enabling the duplication of a Matter node lies in the inability 
to distinguish between devices that share identical credentials. 
To mitigate this issue, credentials should be bound to a specific piece of hardware. 
While identifiers such as MAC and IP addresses may appear as straightforward options, 
Matter mandates their periodic randomization to preserve user privacy, 
rendering them unsuitable for secure identity binding. 
We advocate for the use of more robust and immutable hardware-rooted properties. 
One promising approach is Physical Unclonable Functions (PUFs), 
which produce unique and repeatable responses to cryptographic challenges based on 
intrinsic manufacturing variations~\cite{puf1, puf2}. 
By anchoring credential usage to a device’s PUF-derived identity, 
the credentials become intrinsically non-transferable.
This hardware-rooted binding effectively mitigating MiTM and other impersonation attacks
since the PUF secrets cannot be leaked.

\paragraph{C2 $\rightarrow$ D2\&I1\&I3\&I4: Credential Protection.} 
Another prerequisite of the attack is the compromise of device credentials. 
Naturally, protecting these secrets from unauthorized access becomes a critical line of defense. 
The required level of protection must be elevated, as the threat model assumes 
a physically proximate and well-resourced adversary, 
rather than a purely remote attacker. 
Under such conditions, plaintext credentials residing in flash storage, disk, RAM, or even CPU cache can
be readily extracted using well-known physical attacks~\cite{coldboot,frost,voltboot,untrustzone,recessim}.
Our PoC targets devices that keep credentials in unencrypted flash.
Flash encryption thus arises as an immediate remedy.

We advocate for a combination of software and hardware-based protections. 
From a software perspective, secrets should remain encrypted at rest and only be decrypted in memory 
when strictly necessary for runtime operations. 
Although software retrofit cannot completely eliminate secret leak, the bar is substantially raised since
the attack becomes more sensitive to the timing.
From a hardware standpoint, secure hardware components such as Security Elements (SEs) should 
be adopted where feasible. 
Although IoT devices face cost and resource constraints, 
these must be carefully balanced against the security risks of physical attacks.

\paragraph{C3 $\rightarrow$ D1\&I2\&I3: Periodical Two-way Re-attestation and Disallowing Overriding.} 
Currently, Matter mandates attestation only for devices, not controllers. 
Moreover, credentials used for establishing a CASE session (\eg~NOC) are long-lived; 
once compromised, they can be reused indefinitely to initiate new sessions. 
More concerning, many controllers allow users to enroll devices even if attestation fails.
To mitigate these risks, we advocate for periodic, mutual re-attestation between controllers and devices. 
This would help detect rogue nodes and invalidate outdated or stolen credentials, 
thereby preventing impersonation attacks. 
From an implementation perspective, controllers should enforce attestation 
strictly and disallow the user to override the controller decision
when a device fails attestation.

\paragraph{C4 $\rightarrow$ D3: Network Scanning.}
Prior research has demonstrated that hidden or unauthorized IoT devices can be identified 
through network behavior analysis~\cite{leakypick, snoopdog, lumos}. 
Techniques include fingerprinting MAC-layer traffic~\cite{lumos}, 
correlating user interactions with traffic patterns~\cite{snoopdog}, 
and injecting stimuli to elicit identifiable responses~\cite{leakypick}. 
These passive and active anomaly detection methods can be adapted 
to identify impersonated nodes, even when they possess valid credentials.

\subsection{Real-World Implication}

While our attack was successful on several real products, an estimation on the
number of potentially impacted devices is still unclear given that (1) the
physical protection of the consumer devices can be quite diverse, and (2) the
attacker's expertise, dedication, and resources may vary. Nevertheless, our
small-scale experiment indicates that this threat is real: as long as physical
memory disclosure is achievable, Matter credentials can be quickly extracted and
reused in persistent impersonation attacks. Considering that many IoT gadgets
are budget-friendly, their physical security is usually not a priority, as we
show in this work. Therefore, the potential impact of our attack cannot be
underestimated.

\section{Related Work}\label{Related}

The recent success of Matter has spurred growing interest in its security implications, 
particularly within the IoT and smart home research communities~\cite{ndss-matter, usenix-matter, matter-thesis}. 
Wang et al.~\cite{ndss-matter} analyze the user-facing control interfaces of Matter-enabled devices 
and identify UI design flaws that compromise user control and awareness, 
attributing these issues to vague specifications and inconsistent interpretations by vendors. 
Ma et al.~\cite{usenix-matter} propose a black-box Matter fuzzer that 
leverages large-language models (LLMs) to transform protocol specifications into state machines, 
enabling semantic-aware fuzzing that reveal known and previously undisclosed vulnerabilities. 
Loos et al.~\cite{matter-thesis} investigate theoretical attack vectors stemming from 
specification weaknesses and poor implementation practices, focusing primarily on onboarding threats such as passcode compromise.
Complementing prior analyses, our work provides the first comprehensive security 
evaluation of the Matter standard within shared physical environments. 
We uncover seven fundamental vulnerabilities, demonstrate their real-world exploitability, and outline concrete remediation strategies.

Our work also connects to prior research on multi-user smart homes, where the complex dynamics of shared occupancy give rise to new security and privacy challenges.
Existing systems address vulnerabilities in access delegation, data sharing, and device control in a multi-user or shared domain~\cite{spacelord, chi24, usenix19, teo, wave, rethinking, wan2026characterizing}. 
Wan et al.~\cite{wan2026characterizing} conduct a large-scale empirical study of device access sharing workflows for smart home IoT devices, identifying nine recurring classes of security and privacy risks that arise during the sharing process, including unintended privacy exposure and uncontrolled re-sharing.
Spacelord~\cite{spacelord} enables secure and private time-sharing of smart devices 
in shared spaces by allowing transient users to configure devices with trusted code 
while reverting changes upon their departure to preserve owner control. 
Other efforts seek to balance multi-user control and privacy: 
Zhou et al.~\cite{chi24} propose a polling-based interface to help co-occupants negotiate privacy settings on shared devices,
while Zeng et al.~\cite{usenix19} design a mobile app that manages access controls and 
notifies administrative users of state changes initiated by others. 
TEO~\cite{teo} introduces a co-ownership model that grants transient users control over data they generate, 
backed by a protocol suite for access management, revocation, and group ownership. 
WAVE~\cite{wave} proposes a decentralized authorization framework that 
eliminates reliance on central trust by enabling cryptographically enforced, 
fine-grained delegation and revocation of permissions, 
making it particularly well-suited for dynamic, multi-user IoT environments. 
Our attack shows that in shared or transient environments, Matter’s design can be exploited to achieve persistent, 
fabric-wide control--exposing security risks that existing multi-user access and privacy mechanisms do not address.

Our attack assumes temporary physical access to victim devices, similar to
prior work~\cite{tsang2022fandemic, etrojans, prankscooter,wan2026characterizing}. For example,
Tsang et al.\cite{tsang2022fandemic} induce unintended behavior in IoT
devices by manipulating their power-management circuitry, requiring extensive
reverse engineering and firmware modification. E-Trojans\cite
{etrojans} introduce malicious firmware to Xiaomi e-scooters, enabling
attacks such as undervoltage ransomware and user tracking, while a separate
report~\cite{prankscooter} describes rental e-scooters being tampered with by
replacing their audio files. All of these attacks rely on altering device
firmware or hardware internals. In contrast, although we assume similar
physical access, our attack requires no firmware modification.

Finally, our work fits within the broader smart-home security literature on how interactive and behavioral dynamics create novel threat surfaces~\cite{evasion, interactions, iotmediator, iotmon}. Prior work shows that anomaly detection can be bypassed by manipulating sensor-actuator behaviors~\cite{evasion}, insecure device-cloud-app interactions enable hijacking and firmware theft~\cite{interactions}, control channel conflicts across platforms pose risks in multi-platform homes~\cite{iotmediator}, and inter-app physical interaction chains can produce unsafe unintended behaviors~\cite{iotmon}. Our work advances this literature by revealing protocol-level vulnerabilities in the very mechanisms intended to unify and secure the ecosystem---distinct from the cross-app interactions and behavioral dynamics studied in prior work---and by proposing concrete countermeasures for future smart-home standards.

\section{Conclusions}\label{Conclusions}

This paper highlights critical vulnerabilities in the Matter specification that stem from both design and implementation flaws, exacerbated by the protocol’s open architecture. 
While Matter’s transparency and standardization have accelerated the development of interoperable smart home ecosystems, they have also inadvertently empowered adversaries with sufficient knowledge to 
extract credentials and impersonate legitimate controllers and devices.

We demonstrate that temporary but malicious occupants with physical access to a Matter-based smart home 
can exploit these factors to clone Matter entities and maintain persistent, stealthy control over the smart home environment long after their departure. 
Alarmingly, such adversaries do not need to reverse engineer each target device individually. 
As shown in our evaluation, credentials can be automatically identified, 
and device replicas can be rapidly constructed using firmware templates---facilitated by universal adherence to the Matter specification.

The proposed impersonation attack fundamentally compromises core security properties of smart home systems, 
including confidentiality, integrity, and user privacy, all while remaining largely undetectable. 
Our findings underscore the pressing need to fortify Matter’s security posture beyond its current trust assumptions.
In particular, we advocate for enhanced credential binding, 
mutual authentication, periodic re-attestation, and escalated protections for credential storage and handling.

%

\bibliographystyle{ACM-Reference-Format}
\bibliography{./bibliography}

\appendix

%
%

\section{Ethical Considerations}

We have followed established ethical guidelines for security research, ensuring
that all experiments were conducted in isolated environments without impacting
real-world users or deployed systems. The security flaws, protocol weaknesses,
and implementation concerns presented in this paper were responsibly disclosed
to the Connectivity Standards Alliance (CSA) and the affected manufacturers
prior to submission. The impacted manufacturers were anonymized to avoid
disclosing vulnerabilities publicly. Our goal of conducting and presenting this
research is to improve the overall security posture of the Matter ecosystem and
to support a culture of transparency and collaboration within the community.

\end{document}